\documentclass[12pt]{article}
\usepackage{amsmath}
\usepackage{bm}
\usepackage{color}

\begin{document}
\begin{center}
\begin{large}
{\bf Harmonic oscillator chain in noncommutative phase space with rotational symmetry}
\end{large}
\end{center}

\centerline {Kh. P. Gnatenko \footnote{E-Mail address: khrystyna.gnatenko@gmail.com}}
\medskip
\centerline {\small \it Ivan Franko National University of Lviv, Department for Theoretical Physics,}
\centerline {\small \it 12 Drahomanov St., Lviv, 79005, Ukraine}
\centerline {\small \it  Laboratory for Statistical Physics of Complex Systems}
\centerline {\small \it  Institute for Condensed Matter Physics, NAS of Ukraine, Lviv, 79011, Ukraine}

\begin{abstract}
We consider a quantum space with rotationally invariant noncommutative algebra of coordinates and momenta.  The algebra contains tensors of noncommutativity constructed involving additional coordinates and momenta. In the rotationally invariant noncommutative phase space harmonic oscillator chain is studied.  We obtain that noncommutativity  affects on the frequencies of the system. In the case of a chain of particles with harmonic oscillator interaction we conclude that because of momentum noncommutativity the spectrum of the center-of-mass of the system is discrete and corresponds to the spectrum of harmonic oscillator.

\end{abstract}

\section{Introduction}

Owing to development of String Theory and Quantum Gravity \cite{Witten,Doplicher} studies of idea that space coordinates may be noncommutative has attracted much attention. Noncommutativity of coordinates
leads to existence of  minimal length, minimal area \cite{Romero2003,GnatenkoUFG18}, it leads to space quantization.
Canonical version of noncommutative phase space is characterized by the following algebra
\begin{eqnarray}
[X_{i},X_{j}]=i\hbar\theta_{ij},\label{form101}{}\\{}
[P_{i},P_{j}]=i\hbar\eta_{ij},\label{form1001}{}\\{}
[X_{i},P_{j}]=i\hbar(\delta_{ij}+\gamma_{ij}).\label{form10001}{}
\end{eqnarray}
where $\theta_{ij}$, $\eta_{ij}$, $\gamma_{ij}$ are elements of constant matrixes. Parameters $\gamma_{ij}$ are considered to be defined as $\gamma_{ij}=\sum_k \theta_{ik}\eta_{jk}/4$ \cite{Bertolami}.

Noncommutative algebra (\ref{form101})-(\ref{form10001}) with $\theta_{ij}$, $\eta_{ij}$, $\gamma_{ij}$ being constants is not rotationally invariant \cite{Chaichian,Balachandran1}. Different generalizations of commutation relations (\ref{form101})-(\ref{form10001}) were considered to solve the problem of rotational symmetry breaking in noncommutative space \cite{Moreno,Galikova,Amorim,GnatenkoPLA14}. Many  papers are devoted to studies of position-dependent noncommutativity \cite{Lukierski,Lukierski2009,BorowiecEPL,Borowiec,Borowiec1,Kupriyanov2009,Kupriyanov}, spin noncommutativity \cite{Falomir09,Ferrari13}. The algebras of these types of noncommutativity are rotationally invariant by they are not equivalent to noncommutative algebra of canonical type.

In paper \cite{GnatenkoIJMPA17} a rotationally invariant noncommutative algebra of canonical type was constructed on the basis of idea of generalization of parameters of noncommutativity to a tensors.  Introducing additional coordinates and additional momenta  $\tilde{a}_i$, $\tilde{b}_i$  $\tilde{p}^a_i$, $\tilde{p}^b_i$,  we proposed to define these tensors in the following form
\begin{eqnarray}
\theta_{ij}=\frac{c_{\theta} l^2_{P}}{\hbar}\sum_k\varepsilon_{ijk}\tilde{a}_{k}, \label{form130}\\
\eta_{ij}=\frac{c_{\eta}\hbar}{l^2_{P}}\sum_k\varepsilon_{ijk}\tilde{p}^b_{k}.\label{for130}
\end{eqnarray}
  Constants $c_{\theta}$, $c_{\eta}$ are dimensionless,  $l_P$ is the Planck length. To preserve the rotational symmetry the coordinates and momenta  $\tilde{a}_i$, $\tilde{b}_i$  $\tilde{p}^a_i$, $\tilde{p}^b_i$ are supposed to be governed by a rotationally invariant systems. The systems are considered to be harmonic oscillators
   \begin{eqnarray}
 H^a_{osc}=\hbar\omega_{osc}\left(\frac{(\tilde{p}^{a})^{2}}{2}+\frac{\tilde{a}^{2}}{2}\right),\label{form104}\\
 H^b_{osc}=\hbar\omega_{osc}\left(\frac{(\tilde{p}^{b})^{2}}{2}+\frac{\tilde{b}^{2}}{2}\right),\label{for104}
 \end{eqnarray}
with $\sqrt{{\hbar}}/\sqrt{{m_{osc}\omega_{osc}}}=l_{P}$ and large frequency $\omega_{osc}$ (the distance between energy levels is very large and oscillators are considered to be in the ground states). The algebra for additional coordinates and additional momenta is the following
 \begin{eqnarray}
[\tilde{a}_{i},\tilde{a}_{j}]=[\tilde{b}_{i},\tilde{b}_{j}]=[\tilde{a}_{i},\tilde{b}_{j}]=0,\\{}
[\tilde{p}^{a}_{i},\tilde{p}^{a}_{j}]=[\tilde{p}^{b}_{i},\tilde{p}^{b}_{j}]=[\tilde{p}^{a}_{i},\tilde{p}^{b}_{j}]=0,\\{}
[\tilde{a}_{i},\tilde{p}^{b}_{j}]=[\tilde{b}_{i},\tilde{p}^{a}_{j}]=0,\\{}
[\tilde{a}_{i},X_{j}]=[\tilde{a}_{i},P_{j}]=[\tilde{p}^{b}_{i},X_{j}]=[\tilde{p}^{b}_{i},P_{j}]=0,\\{}
[\tilde{a}_{i},\tilde{p}^{a}_{j}]=[\tilde{b}_{i},\tilde{p}^{b}_{j}]=i\delta_{ij}.
 \end{eqnarray}
 Therefore, we have  $[\theta_{ij}, X_k]=[\theta_{ij}, P_k]=[\eta_{ij}, X_k]=[\eta_{ij}, P_k]=[\gamma_{ij}, X_k]=[\gamma_{ij}, P_k]=0$ as in the case of canonical noncommutativity (\ref{form101})-(\ref{form10001}) with $\theta_{ij}$, $\eta_{ij}$, $\gamma_{ij}$ being constants.

 In the present paper we study influence of noncommutativity of coordinates and noncommutativity of momenta on the spectrum of a harmonic oscillator chain.
Studies of a system of harmonic oscillators are important in various fields of physics among them molecular spectroscopy and quantum chemistry \cite{Ikeda,Fillaux,Hong90,Michelot92}, quantum optics \cite{Caves85,Schumaker85,Plenio}, nuclei physics \cite{Isgur,Glozman,Capstick}, quantum information processing \cite{Audenaert,Plenio,Plenio1}.

 Harmonic oscillator was intensively studied in the frame of noncommutative algebra \cite{Hatzinikitas,Kijanka,Jing,Smailagic,Smailagic1,Muthukumar,Alvarez,Djemai,Dadic,Giri,Geloun,Abreu,Saha11,Nath,Shyiko}. Recently experiments with micro- and nano-oscillators were implemented for probing  minimal length \cite{Bawaj}. In noncommutative space of canonical type   two coupled harmonic oscillators were studied in \cite{Jellal,Bing,GnatenkoJPS17}. In \cite{Gnatenko_arx181} a spectrum of a system of $N$ oscillators  interacting with each other (symmetric network of coupled harmonic oscillators) has been examined in rotationally invariant noncommutative phase space.  In \cite{Daszkiewicz} classical $N$ interacting harmonic oscillators were examined in noncommutative space-time.  In \cite{BastosPhysA,Laba} influence of noncommutativity of coordinates and noncommutativity of momenta on the properties of a system of free particles was examined.

The paper is organized as follows.  In Section 2 we study energy levels of a harmonic oscillator chain in rotationally invariant noncommutative phase space. Particular case of a chain of particles with harmonic oscillator interaction is  examined.  Conclusions are presented in Section 3.

\section{Spectrum of harmonic oscillator chain in rotationally invariant noncommutative phase space}

Let us consider a chain of $N$ interacting harmonic oscillators with masses $m$ and frequencies $\omega$ in a space with (\ref{form101})-(\ref{form10001}) and  (\ref{form130}), (\ref{for130}) in the case of the closed configuration of the system. So, let us study the following Hamiltonian
\begin{eqnarray}
 H_s=\sum_{n=1}^N\frac{( {\bf P}^{(n)})^{2}}{2m}+\sum_{n=1}^N\frac{m\omega^2( {\bf X}^{(n)})^{2}}{2}+\nonumber\\+k{\sum_{n=1}^N}({\bf X}^{(n+1)}-{\bf X}^{(n)})^2\label{form777}
\end{eqnarray}
with periodic boundary conditions ${\bf X}^{(N+1)}={\bf X}^{(1)}$, $k$ is a constant.

In general case coordinates and momenta which correspond to different particles satisfy noncommutative algebra with different tensors of noncommutativity. We have
 \begin{eqnarray}
[X^{(n)}_{i},X^{(m)}_{j}]=i\hbar\delta_{mn}\theta^{(n)}_{ij},\label{ffor101}\\{}
[X^{(n)}_{i},P^{(m)}_{j}]=i\hbar\delta_{mn}\left(\delta_{ij}+\sum_k\frac{\theta^{(n)}_{ik}\eta^{(m)}_{jk}}{4}\right),\label{for1001}\\{}
[P^{(n)}_{i},P^{(m)}_{j}]=i\hbar\delta_{mn}\eta^{(n)}_{ij},\label{ffor10001}\\{}
\theta^{(n)}_{ij}=\frac{c_{\theta}^{(n)}l_P^2}{\hbar}\sum_k\varepsilon_{ijk}\tilde{a}_{k}, \label{tn}\\
\eta^{(n)}_{ij}=\frac{c_{\eta}^{(n)}\hbar}{l_P^2}\sum_k\varepsilon_{ijk}\tilde{p}^b_{k},\label{etn}
 \end{eqnarray}
where indexes $m,n=(1...N)$ label the particles \cite{GnatenkoIJMPA18}.

Because of presence of additional coordinates and momenta in (\ref{tn}), (\ref{etn}) we have to study Hamiltonian which include Hamiltonians of harmonic oscillators
\begin{eqnarray}
H=H_s+H^a_{osc}+H^b_{osc}\label{total}
\end{eqnarray}
The noncommutative coordinates and noncommutative momenta can be represented as
\begin{eqnarray}
X^{(n)}_{i}=x^{(n)}_{i}+\frac{1}{2}[{\bm \theta}^{(n)}\times{\bf p}^{(n)}]_i,\label{repx0}\\
P^{(n)}_{i}=p^{(n)}_{i}-\frac{1}{2}[{\bf x}^{(n)}\times{\bm \eta}^{(n)}]_i,\label{repp0}
\end{eqnarray}
where coordinates and momenta $x^{(n)}_i$, $p^{(n)}_i$ satisfy  the ordinary commutation relations
 \begin{eqnarray}
[x^{(n)}_{i},x^{(m)}_{j}]=[p^{(n)}_{i},p^{(m)}_{j}]=0,\label{orx}\\{}
[x^{(n)}_{i},p^{(m)}_{j}]=i\hbar\delta_{mn}.\label{orp}
 \end{eqnarray}
 and vectors ${\bm \theta}^{(n)}$, ${\bm \eta}^{(n)}$ have the components
$\theta^{(n)}_i=\sum_{jk}\varepsilon_{ijk}{\theta^{n}_{jk}}/2,$
$\eta^{(n)}_i=\sum_{jk}\varepsilon_{ijk}{\eta^{(n)}_{jk}}/2.$
In our paper \cite{GnatenkoIJMPA18} we proposed the constants $c^{(n)}_{\theta}$, $c^{(n)}_{\eta}$ in tensors of noncommutativity to be determined by mass as $c^{(n)}_{\theta}m_n=\tilde{\gamma}=const$, $c^{(n)}_{\eta}/{m_n}=\tilde{\alpha}=const$ with $\tilde{\gamma}$, $\tilde{\alpha}$ being the same for different particles. Therefore one has
 \begin{eqnarray}
\theta^{(n)}_{ij}=\frac{\tilde{\gamma}l_P^2}{m_n\hbar}\sum_k\varepsilon_{ijk}\tilde{a}_{k}, \label{ctn}\\
\eta^{(n)}_{ij}=\frac{\tilde{\alpha} \hbar m_n}{l_P^2}\sum_k\varepsilon_{ijk}\tilde{p}^b_{k}.\label{cetn}
 \end{eqnarray}
Determination of tensors of noncommutativity in forms (\ref{ctn}), (\ref{cetn}) gives a possibility to consider noncommutative coordinates as kinematic variables \cite{GnatenkoIJMPA18}, to recover the weak equivalence principle \cite{Gnatenko_arxiv}.
Taking into account (\ref{ctn}), (\ref{cetn}), in the case when the system consists of oscillators with the same masses one has $\theta^{(n)}_{ij}=\theta_{ij}$, $\eta^{(n)}_{ij}=\eta_{ij}$.
Using (\ref{repx0})-(\ref{repp0}) the Hamiltonian $H_s$ reads
\begin{eqnarray}
 H_s=\sum_{n=1}^N\left(\frac{({\bf p}^{(n)})^{2}}{2m}+\frac{m\omega^2( {\bf x}^{(n)})^{2}}{2}+\right.\nonumber\\ \left.+k({\bf x}^{(n+1)}-{\bf x}^{(n)})^2-\frac{({\bm \eta}\cdot[{\bf x}^{(n)}\times{\bf p}^{(n)}])}{2m}-\right.\nonumber\\ \left.-\frac{m\omega^2({\bm \theta}\cdot[{\bf x}^{(n)}\times{\bf p}^{(n)}])}{2}-\right.\nonumber\\\left.-
 {k}({\bm \theta}\cdot[({{\bf x}}^{(n+1)}-{{\bf x}}^{(n)})\times ({\bf p}^{(n+1)}-{\bf p}^{(n)})])+\right.\nonumber\\\left.+\frac{[{\bm \eta}\times{\bf x}^{(n)}]^2}{8m}+\frac{m\omega^2}{8}[{\bm \theta}\times{\bf p}^{(n)}]^2+\right.\nonumber\\\left.+\frac{k}{4}[{\bm \theta}\times ({\bf p}^{(n+1)}-{\bf p}^{(n)})]^2\right).\label{orm777}
\end{eqnarray}

In  \cite{GnatenkoIJMPA18} we showed that up to the second order in $\Delta H$  defined as
\begin{eqnarray}
\Delta H=H_s-\langle H_s\rangle_{ab},
\end{eqnarray}
Hamiltonian
\begin{eqnarray}
H_0=\langle H_s\rangle_{ab}+H^a_{osc}+H^b_{osc},\label{2h0}
\end{eqnarray}
can be studied because up to the second order in the perturbation theory the corrections to spectrum of $H_0$ caused by terms $\Delta H=H-H_0=H_s-\langle H_s\rangle_{ab}$ vanish.
Here notation $\langle...\rangle_{ab}$ is used for averaging over the eigenstates of  $H^a_{osc}$ $H^b_{osc}$ which are well known
$\langle...\rangle_{ab}=\langle\psi^{a}_{0,0,0}\psi^{b}_{0,0,0}|...|\psi^{a}_{0,0,0}\psi^{b}_{0,0,0}\rangle$.
For the harmonic oscillator chain we have
\begin{eqnarray}
\Delta H=\sum_{n=1}^N\left(\frac{[{\bm \eta}\times{\bf x}^{(n)}]^2}{8m}+\frac{m\omega^2}{8}[{\bm \theta}\times{\bf p}^{(n)}]^2-\right.\nonumber\\\left.-\frac{m\omega^2({\bm \theta}\cdot[{\bf x}^{(n)}\times{\bf p}^{(n)}])}{2}-\frac{({\bm \eta}\cdot[{\bf x}^{(n)}\times{\bf p}^{(n)}])}{2m}-\right.\nonumber\\\left.-
{k}{\bm \theta}\cdot[({{\bf x}}^{(n+1)}-{{\bf x}}^{(n)})\times ({\bf p}^{(n+1)}-{\bf p}^{(n+1)})]+\right.\nonumber\\\left.+\frac{k}{4}[{\bm \theta}\times ({\bf p}^{(n+1)}-{\bf p}^{(n)})]^2-\frac{\langle\eta^2\rangle({\bf x}^{(n)})^2}{12m}-\right.\nonumber\\\left.-\frac{\langle\theta^2\rangle m\omega^2({\bf p}^{(n)})^2}{12}-\frac{k}{6}\langle{\theta}^2\rangle ({\bf p}^{(n+1)}-{\bf p}^{(n)})^2\right). \nonumber\\\label{delta}
 \end{eqnarray}
here we take into account that $\langle\psi^{a}_{0,0,0}|\tilde{a}_i|\psi^{a}_{0,0,0}\rangle=\langle\psi^{b}_{0,0,0}|\tilde{p}_i|\psi^{b}_{0,0,0}\rangle=0$ and use the following notations
\begin{eqnarray}
\langle\theta_i\theta_j\rangle=\nonumber\\=\frac{c_{\theta}^2l_P^4}{\hbar^2}\langle\psi^{a}_{0,0,0}| \tilde{a}_i\tilde{a}_j|\psi^{a}_{0,0,0}\rangle=\frac{c_{\theta}^2l_P^4}{2\hbar^2}\delta_{ij}=\frac{\langle\theta^2\rangle\delta_{ij}}{3},\label{thetar2}\\
\langle\eta_i\eta_j\rangle=\nonumber\\= \frac{\hbar^2 c_{\eta}^2}{l_P^4}\langle\psi^{b}_{0,0,0}| \tilde{p}^{b}_i\tilde{p}^{b}_j|\psi^{b}_{0,0,0}\rangle=\frac{\hbar^2 c_{\eta}^2}{2 l_P^4}\delta_{ij}=\frac{\langle\eta^2\rangle\delta_{ij}}{3}.\label{etar2}
\end{eqnarray}

 So, analyzing  the form of $\Delta H$ (\ref{delta}), we have that up to the second order in the parameters of noncommutativity one can study Hamiltonian $H_0$. This Hamiltonian for convenience can be rewritten as

 \begin{eqnarray}
H_0=\sum_{n=1}^N\left(\frac{({\bf p}^{(n)})^{2}}{2m_{eff}}+\frac{m_{eff}\omega_{eff}^2( {\bf x}^{(n)})^{2}}{2}+\right.\nonumber\\\left.+k({{\bf x}}^{(n+1)}-{{\bf x}}^{(n)})^2+\right.\nonumber\\ \left.+\frac{k}{6}\langle{\theta}^2\rangle ({\bf p}^{(n+1)}-{\bf p}^{(n)})^2+H^a_{osc}+H^b_{osc} \right),\label{h5}
  \end{eqnarray}
with
 \begin{eqnarray}
 m_{eff}={m}\left({1+\frac{m^2\omega^2\langle\theta^2\rangle}{6}}\right)^{-1},\label{meff}\\
 \omega_{eff}=\left({\omega^2+\frac{\langle\eta^2\rangle}{6m^2}}\right)^{\frac{1}{2}}\left({1+\frac{m^2\omega^2\langle\theta^2\rangle}{6}}\right)^{\frac{1}{2}}.\label{omegaeff}
   \end{eqnarray}
The terms  $H^a_{osc}+H^b_{osc}$ commute with $H_0$. Coordinates and momenta ${\bf x}^{(n)}$, ${\bf p}^{(n)}$ satisfy (\ref{orx}), (\ref{orp}). Let us rewrite $H_0$ as
\begin{eqnarray}
H_0=\nonumber\\\frac{\hbar\omega_{eff}}{2}\sum_{n}\left(1+\frac{4km_{eff}\langle{\theta}^2\rangle}{3}\sin^2\frac{\pi n}{N}\right){\tilde {\bf p}}^{(n)}({\tilde{\bf  p}^{(n)}})^{\dag}+\nonumber\\+\frac{\hbar\omega^2_{eff}}{2}\sum_{n}\left(1+\frac{8k}{m_{eff}\omega^2_{eff}}\sin^2\frac{\pi n}{N}\right){\tilde {\bf x}^{(n)}}({\tilde {\bf x}^{(n)}})^{\dag},\nonumber\\
\end{eqnarray}
using
 \begin{eqnarray}
{\bf x}^{(n)}=\sqrt{\frac{\hbar}{Nm_{eff}\omega_{eff}}}\sum^{N}_{l=1}\exp\left({\frac{2\pi i nl}{N}}\right)\tilde{{\bf x}}^{(l)},\\
{\bf p}^{(n)}=\sqrt{\frac{\hbar m_{eff}\omega_{eff}}{N}}\sum^{N}_{l=1}\exp\left(-{\frac{2\pi i nl}{N}}\right)\tilde{{\bf p}}^{(l)}
   \end{eqnarray}
(see, for example, \cite{Plenio}). Introducing operators $a^{(n)}_j$ defined as
 \begin{eqnarray}
a^{(n)}_j=\frac{1}{\sqrt{2w_n}}\left(w_n\tilde{x}^{(n)}_j+i\tilde{p}^{(n)}_j\right),\\
w_n=\left(1+\frac{8k}{m_{eff}\omega^2_{eff}}\sin^2\frac{\pi n}{N}\right)^{\frac{1}{2}}\times\nonumber\\\times\left(1+\frac{4km_{eff}\langle{\theta}^2\rangle}{3}\sin^2\frac{\pi n}{N}\right)^{-\frac{1}{2}}
   \end{eqnarray}
we have
\begin{eqnarray}
H_0=\hbar\omega_{eff}\sum^N_{n=1}\sum^{3}_{j=1}\left(1+\frac{4km_{eff}\langle{\theta}^2\rangle}{3}\sin^2\frac{\pi n}{N}\right)^{\frac{1}{2}}\times\nonumber\\\times\left(1+\frac{8k}{m_{eff}\omega^2_{eff}}\sin^2\frac{\pi n}{N}\right)^{\frac{1}{2}}\left((a^{(n)}_j)^{\dag} a^{(n)}_j+\frac{1}{2}\right).\nonumber\\
\end{eqnarray}
The spectrum of $H_0$ reads
\begin{eqnarray}
E_{\{n_1\},\{n_2\},\{n_3\}}=\hbar\sum^N_{a=1}\left(\omega^2_{eff}+\frac{8k}{m_{eff}}\sin^2\frac{\pi a}{N}\right)^{\frac{1}{2}}\times\nonumber\\\times\left(1+\frac{4km_{eff}\langle{\theta}^2\rangle}{3}\sin^2\frac{\pi a}{N}\right)^{\frac{1}{2}}\left(n^{(a)}_1+n^{(a)}_2+\right.\nonumber\\\left.+n^{(a)}_3+\frac{3}{2}\right)=
\sum^N_{a=1}\hbar\omega_a\left(n^{(a)}_1+n^{(a)}_2+n^{(a)}_3+\frac{3}{2}\right),\nonumber\\\label{ennh}
\end{eqnarray}
where $n^{(a)}_i$ are quantum numbers ($n^{(a)}_i=0,1,2...$). Taking into account (\ref{meff}), (\ref{omegaeff}) the frequencies reads

\begin{eqnarray}
\omega^2_a=\left(\omega^2+\frac{\langle\eta^2\rangle}{6m^2}\right)\left(1+\frac{m^2\omega^2\langle\theta^2\rangle}{6}+\right.\nonumber\\\left.+
\frac{4k^2m\langle \theta^2\rangle}{3}\sin^2\frac{\pi a}{N}\right)+\frac{8k}{m}\sin^2\frac{\pi a}{N}+\nonumber\\+\frac{32k^2\langle\theta^2\rangle}{3}\sin^4\frac{\pi a}{N}.\label{omaa}
\end{eqnarray}

For a chain of particles with harmonic oscillator interaction, describing by Hamiltonian (\ref{form777}) with $\omega=0$, up to the second order in the parameters of noncommutativity one has

\begin{eqnarray}
E_{\{n_1\},\{n_2\},\{n_3\}}=\nonumber\\=
\sum^N_{a=1}\hbar\omega_a\left(n^{(a)}_1+n^{(a)}_2+n^{(a)}_3+\frac{3}{2}\right),\label{enh}
\end{eqnarray}
with
\begin{eqnarray}
\omega^2_a=\frac{8k}{m}\sin^2\frac{\pi a}{N}+\frac{\langle\eta^2\rangle}{6m^2}+ \frac{32k^2\langle{\theta}^2\rangle}{3}\sin^4\frac{\pi a}{N}.\label{omegaa}
\end{eqnarray}
It is worth noting that in the case of a space with noncommutative coordinates and commutative momenta  (\ref{form101})-(\ref{form10001}) with (\ref{form130}) and $\eta_{ij}=0$  the spectrum of a chain of particles with harmonic oscillator reads (\ref{enh}) with
\begin{eqnarray}
\omega^2_a=\frac{8k}{m}\sin^2\frac{\pi a}{N}+\frac{32k^2\langle{\theta}^2\rangle}{3}\sin^4\frac{\pi a}{N}.\label{omegat}
\end{eqnarray}
Note that $\omega^2_N=0$ and corresponds to the spectrum of the center-of-mass of the system. Noncommutativity of momenta leads to discrete spectrum of the center-of-mass of a chain of interacting particles. From (\ref{enh}), (\ref{omegaa}) we have that the spectrum of the center-of-mass of the system corresponds to the spectrum of three dimensional harmonic oscillator with frequency determined as
\begin{eqnarray}
\omega^2_N=\frac{\langle\eta^2\rangle}{6m^2}.\label{omet}
\end{eqnarray}

In the limit $\langle\theta^2\rangle\rightarrow0$, $\langle\eta^2\rangle\rightarrow0$ form (\ref{omaa}) we obtain well known result $\omega^2_a=\omega^2+\frac{8k}{m}\sin^2\frac{\pi a}{N}$.

\section{Conclusions}

Rotationally invariant algebra with noncommutativity of coordinates and noncommutativity of momenta has been considered. The algebra is constructed involving additional coordinates and additional momenta  (\ref{form101})-(\ref{form10001}) with (\ref{form130}), (\ref{for130}).   We have studied influence of noncommutativity on the spectrum of harmonic oscillator chain with periodic boundary conditions. For this purpose the total Hamiltonian has been examined (\ref{total}) and energy levels of harmonic oscillator chain have been obtained up to the second order in the parameters of noncommutativity.  We have found that noncommutativity does not change the form of the chain's spectrum (\ref{ennh}). Noncommutativity of coordinates and noncommutativity of momenta affects on the frequencies of the system (\ref{omaa}).

The case of a chain of particles with harmonic oscillator interaction describing by Hamiltonian (\ref{form777}) with $\omega=0$ has been studied.  We have obtained that the spectrum of the center-of-mass of the system is discrete because of noncommutativity of momenta. This spectrum corresponds to the spectrum of harmonic oscillator with frequency (\ref{omet}).

\vskip3mm \textit{Acknowledgement.}
The author thanks Prof. V. M. Tkachuk for his
advices and  support during research studies. The publication contains the results of studies conducted by President's of Ukraine grant for competitive projects (F-75).


\begin{thebibliography}{99}

\bibitem{Witten} N. Seiberg, E. Witten. String theory and noncommutative geometry. {\it J. High Energy Phys.} {\bf 9909}, 032 (1999).
\bibitem{Doplicher} S. Doplicher, K. Fredenhagen, J.E. Roberts. Spacetime quantization induced by classical gravity. {\it Phys. Lett. B} {\bf 331}, 39 (1994). doi: 10.1016/0370-2693(94)90940-7
\bibitem{Romero2003} J.M. Romero, J.A. Santiago, J.D. Vergara. Note about the quantum of area in a noncommutative space. {\it Phys. Rev. D} {\bf 68}, 067503 (2003). doi: 10.1103/PhysRevD.68.067503
\bibitem{GnatenkoUFG18} Kh. P. Gnatenko, V. M. Tkachuk. Lenght in a noncommutative phase space. {\it Ukr. J. Phys.} {\bf 63}, 102, (2018) doi: 10.15407/ujpe63.2.102.

\bibitem{Bertolami} O. Bertolami, R. Queiroz. Phase-space noncommutativity and the Dirac equation. {\it Phys. Lett. A} {\bf 375}, 4116 (2011). doi: 10.1016/j.physleta.2011.09.053
\bibitem{Chaichian}M. Chaichian, M.M. Sheikh-Jabbari, A. Tureanu. Hydrogen atom spectrum and the lamb shift in noncommutative QED. {\it Phys. Rev. Lett.} 86, 2716 (2001). doi: 10.1103/PhysRevLett.86.2716

\bibitem{Balachandran1} A. P. Balachandran, P. Padmanabhan. Non-Pauli effects from noncommutative spacetimes. {\it J. High Energy Phys.} {\bf 1012}, 001 (2010).

\bibitem{Moreno} E. F. Moreno. Spherically symmetric monopoles in noncommutative space. {\it Phys. Rev. D} {\bf 72}, 045001 (2005). doi: 10.1103/PhysRevD.72.045001
\bibitem{Galikova} V. G\'alikov\'a, P. Presnajder. Hydrogen atom in fuzzy spaces-Exact solution. {\it J. Phys: Conf. Ser.} {\bf 343}, 012096 (2012). doi:10.1088/1742-6596/343/1/012096
\bibitem{Amorim} R. Amorim. Tensor operators in noncommutative quantum mechanics. {\it Phys. Rev. Lett.} {\bf 101}, 081602 (2008). doi: 10.1103/PhysRevLett.101.081602
\bibitem{GnatenkoPLA14} Kh.P. Gnatenko, V. M. Tkachuk. Hydrogen atom in rotationally invariant noncommutative space. {\it Phys. Lett. A} {\bf 378}, 3509 (2014). doi: 10.1016/j.physleta.2014.10.021
\bibitem{Lukierski} M. Daszkiewicz, J. Lukierski, M. Woronowicz, Towards quantum noncommutative -deformed field theory {\it Phys. Rev. D} {\bf77}, 105007 (2008) doi: 10.1103/PhysRevD.77.105007
\bibitem{Lukierski2009} M. Daszkiewicz, J. Lukierski, M. Woronowicz. $\kappa$-deformed oscillators, the choice of star product and free $\kappa$-deformed quantum fields {\it J. Phys. A: Math. Theor.} {\bf42}, 355201 (2009). doi: 10.1088/1751-8113/42/35/355201
\bibitem{BorowiecEPL} A. Borowiec, Kumar S. Gupta, S. Meljanac,  A. Pachol, Constraints on the quantum gravity scale from $\kappa$-Minkowski spacetime {\it EPL } {\bf92}, 20006 (2010). doi: 10.1209/0295-5075/92/20006
\bibitem{Borowiec} A. Borowiec, J. Lukierski, A. Pachol.
Twisting and-Poincare. {\it J. Phys. A: Math. Theor.} {\bf47} 405203 (2014). doi: 10.1088/1751-8113/47/40/405203
\bibitem{Borowiec1} A. Borowiec, A. Pachol. $\kappa$ Deformations and Extended $\kappa$-Minkowski Spacetimes {\it SIGMA } {\bf10}, 107 (2014). doi: 10.3842/SIGMA.2014.107
\bibitem{Kupriyanov2009}  M. Gomes, V.G. Kupriyanov. Position-dependent noncommutativity in quantum mechanics. {\it Phys. Rev. D} {\bf79}, 125011 (2009). doi: 10.1103/PhysRevD.79.125011
\bibitem{Kupriyanov} V. G. Kupriyanov. A hydrogen atom on curved noncommutative space. {\it J. Phys. A: Math. Theor.} {\bf 46}, 245303 (2013). doi: 10.1088/1751-8113/46/24/245303

\bibitem{Falomir09} H. Falomir, J. Gamboa, J. López-Sarrión, F. Mendez, P.A.G. Pisani. Magnetic-dipole spin effects in noncommutative quantum mechanics {\it Phys. Lett. B}
{\bf680}  384 (2009). doi: 10.1016/j.physletb.2009.09.007
\bibitem{Ferrari13} A.F. Ferrari, M. Gomes, V.G. Kupriyanov, C.A. Stechhahn, Dynamics of a Dirac fermion in the presence of spin noncommutativity, {\it Phys. Lett. B} {\bf718}, 1475 (2013). doi: 10.1016/j.physletb.2012.12.010



\bibitem{GnatenkoIJMPA17} Kh. P. Gnatenko, V. M. Tkachuk. Noncommutative phase space with rotational symmetry and hydrogen atom. {\it Int. J. Mod. Phys. A} {\bf32},  1750161  (2017). doi: 10.1142/S0217751X17501615



\bibitem{Ikeda} S. Ikeda, F. Fillaux. Incoherent elastic-neutron-scattering study of the vibrational dynamics and spin-related symmetry of protons in the  $KHCO_3$ crystal. {\it Phys. Rev. B} {\bf 59} 4134 (1999). doi: 10.1103/PhysRevB.59.4134
\bibitem{Fillaux}  F. Fillaux. Quantum entanglement and nonlocal proton transfer dynamics in dimers of formic acid and analogues. {\it Chem. Phys. Lett.} {\bf 408} 302 (2005).doi: 10.1016/j.cplett.2005.04.069
\bibitem{Hong90} Fan Hong-yi. Unitary transformation for four Harmonically coupled identical oscillators {\it Phys. Rev. A}, {\bf42}, 4377 (1990). doi: 10.1103/PhysRevA.42.4377
\bibitem{Michelot92}  F. Michelot. Solution for an arbitrary number of coupled identical oscillators. {\it Phys. Rev. A} {\bf45}, 4271 (1992). doi: 10.1103/PhysRevA.45.4271


\bibitem{Caves85} C.M. Caves, B.L. Schumaker, New formalism for two-photon quantum optics. I. Quadrature phases and squeezed states {\it Phys. Rev. A} {\bf 31}, 3068 (1985). doi: 10.1103/PhysRevA.31.3068
\bibitem{Schumaker85}  B.L. Schumaker, C.M. Caves. New formalism for two-photon quantum optics. II. Mathematical foundation and compact notation. {\it Phys. Rev. A} {\bf 31}, 3093 (1985). doi: 10.1103/PhysRevA.31.3093

\bibitem{Plenio}   M.B. Plenio, J. Hartley, J. Eisert, Dynamics and manipulation of entanglement in coupled harmonic systems with many degrees of freedom {\it New J. Phys.} {\bf6}, 36 (2004). doi: 10.1088/1367-2630/6/1/036

\bibitem{Isgur} N. Isgur, G. Karl. P-wave baryons in the quark model. {\it Phys. Rev. D} {\bf 18}, 4187 (1978). doi: 10.1103/PhysRevD.18.4187
\bibitem{Glozman} L. Ya. Glozman, D.O. Riska. The Spectrum of the nucleons and the strange hyperons and chiral dynamics. {\it Phys. Rept.} {\bf 268} 263, (1996) doi: 	10.1016/0370-1573(95)00062-3
\bibitem{Capstick} S. Capstick, W. Roberts. Quark models of baryon masses and decays. {\it Prog. Part. Nucl. Phys.} {\bf 45}, 241, (2000). doi: 10.1016/S0146-6410(00)00109-5


\bibitem{Audenaert} K. Audenaert, J. Eisert, M.B. Plenio, R.F. Werner.  Entanglement properties of the harmonic chain. {\it Phys. Rev. A} {\bf66}, 042327 (2002). doi: 10.1103/PhysRevA.66.042327
\bibitem{Plenio1}  M. B Plenio, F. L Semiao. High efficiency transfer of quantum information and multiparticle entanglement generation in translation-invariant quantum chains. {\it New J. Phys.} {\bf7}, 73 (2005).





\bibitem{Hatzinikitas} A. Hatzinikitas, I. Smyrnakis. The noncommutative harmonic oscillator in more than one dimension. {\it J. Math. Phys.} {\bf43},  113 (2002). doi: 10.1063/1.1416196
\bibitem{Kijanka} A. Kijanka, P. Kosinski. Noncommutative isotropic harmonic oscillator. {\it Phys. Rev. D} {\bf70},  127702 (2004). doi: 10.1103/PhysRevD.70.127702
\bibitem{Jing} Jing Jian, Jian-Feng Chen. Non-commutative harmonic oscillator in magnetic field and continuous limit. {\it Eur. Phys. J. C} {\bf60}, 669 (2009). doi: 10.1140/epjc/s10052-009-0950-1
\bibitem{Smailagic} A. Smailagic, E. Spallucci. Isotropic representation of the noncommutative 2D harmonic oscillator. {\it Phys. Rev. D} {\bf65}, 107701 (2002). doi: 10.1103/PhysRevD.65.107701
\bibitem{Smailagic1} A. Smailagic, E. Spallucci. Noncommutative 3D harmonic oscillator. {\it J. Phys. A} {\bf35}, 363 (2002). doi: 10.1088/0305-4470/35/26/103
\bibitem{Muthukumar} B. Muthukumar, P. Mitra. Noncommutative oscillators and the commutative limit. {\it Phys. Rev. D} {\bf66}  027701 (2002). doi: 10.1103/PhysRevD.66.027701
\bibitem{Alvarez} P. D. Alvarez, J. Gomis, K. Kamimura, M. S. Plyushchay. Anisotropic harmonic oscillator, non-commutative Landau problem and exotic Newton–Hooke symmetry. {\it Phys. Lett. B} {\bf 659}  906 (2008). doi: 10.1016/j.physletb.2007.12.016
\bibitem{Djemai} A. E. F. Djemai, H. Smail. On quantum mechanics on noncommutative quantum phase space. {\it Commun. Theor. Phys.} {\bf41}, 837 (2004). doi: 10.1088/0253-6102/41/6/837
\bibitem{Dadic} I. Dadic, L. Jonke, S. Meljanac. Harmonic oscillator on noncommutative spaces. {\it Acta  Phys. Slov.} {\bf 55}  149 (2005).
\bibitem{Giri}  P. R. Giri, P. Roy. The non-commutative oscillator, symmetry and the Landau problem. {\it Eur. Phys. J. C} {\bf57}, 835 (2008). doi: 10.1140/epjc/s10052-008-0705-4
\bibitem{Geloun} J. Ben Geloun,  S. Gangopadhyay, F. G. Scholtz. Harmonic oscillator in a background magnetic field in noncommutative quantum phase-space. {\it EPL} {\bf86}, 51001 (2009). doi: 10.1209/0295-5075/86/51001
\bibitem{Abreu}  E. M.C. Abreu, M. V. Marcial, A. C.R. Mendes,  W. Oliveira. Analytical and numerical analysis of a rotational invariant D= 2 harmonic oscillator in the light of different noncommutative phase-space configurations. {\it JHEP} 2013, 138 (2013). doi: 10.1007/JHEP11(2013)138
\bibitem{Saha11}  A. Saha, S. Gangopadhyay, S. Saha. Noncommutative quantum mechanics of a harmonic oscillator under linearized gravitational waves. {\it Phys. Rev. D} {\bf83}, 025004 (2011). doi: 10.1103/PhysRevD.83.025004
\bibitem{Nath} D. Nath, P. Roy. Noncommutative anisotropic oscillator in a homogeneous magnetic field. {\it Ann. Phys}, {\bf377},  115  (2017). doi: 10.1016/j.aop.2016.12.028
\bibitem{Shyiko} Kh. P. Gnatenko, O. V. Shyiko. Effect of noncommutativity on the spectrum of free particle and harmonic oscillator in rotationally invariant noncommutative phase space. {\it Mod. Phys. Lett. A} {\bf33},  1850091  (2018). doi: 10.1142/S0217732318500918


\bibitem{Bawaj} M. Bawaj, et. al. Probing deformed commutators with macroscopic harmonic oscillators. {\it Nature Commun.} {\bf6}, 7503 (2015). doi: 10.1038/ncomms8503



\bibitem{Jellal} A. Jellal, El Hassan El Kinani, M. Schreiber. Two coupled harmonic oscillators on noncommutative plane. {\it Int. J. Mod. Phys. A} {\bf20}, 1515 (2005). doi: 10.1142/S0217751X05020835
\bibitem{Bing} Bing-Sheng Lin, Si-Cong Jing, Tai-Hua Heng. Deformation quantization for coupled harmonic oscillators on a general noncommutative space {\it Mod. Phys. Lett. A} {\bf 23}, 445, (2008). doi: 10.1142/S0217732308023992
\bibitem{GnatenkoJPS17}  Kh. P. Gnatenko, V. M. Tkachuk. Two-particle system with harmonic oscillator interaction in noncommutative phase space {\it J. Phys. Stud.} {\bf 21},  3001  (2017).
\bibitem{Gnatenko_arx181} Kh. P. Gnatenko. System of interacting harmonic oscillators in rotationally invariant noncommutative phase space (2018) arXiv:1808.08515


\bibitem{Daszkiewicz} M.~Daszkiewicz, C.J.~Walczyk. Classical Mechanics of Many Particles Defined on Canonically Deformed Nonrelativistic Spacetime. {\it Mod. Phys. Lett} A {\bf26}, 819 (2011). doi: 10.1142/S0217732311035328
\bibitem{BastosPhysA} C. Bastos, A. E. Bernardini, J. F. G. Santos. Probing phase-space noncommutativity through quantum mechanics and thermodynamics of free particles and quantum rotors. {\it Physica A} {\bf438},  340 (2015). doi: 10.1016/j.physa.2015.07.009
\bibitem{Laba} Kh. P. Gnatenko, H. P. Laba, V. M. Tkachuk. Features of free particles system motion in noncommutative phase space and conservation of the total momentum. {\it Mod. Phys. Lett. A} {\bf33},  1850131 (2018). doi: 10.1142/S0217732318501316




\bibitem{GnatenkoIJMPA18} Kh. P. Gnatenko, V. M. Tkachuk. Composite system in rotationally invariant noncommutative phase space. {\it Int. J. Mod. Phys. A {\bf 33}},  1850037  (2018). doi: 10.1142/S0217751X18500379
\bibitem{Gnatenko_arxiv} Kh. P. Gnatenko. Rotationally invariant noncommutative phase space of canonical type with recovered weak equivalence principle. (2018) arXiv:1808.00498.

\bibitem{GnatenkoPLA13} Kh. P. Gnatenko.
Composite system in noncommutative space and the equivalence principle {\it Phys. Lett. A} {\bf 377}, 3061 (2013). doi: 10.1016/j.physleta.2013.09.036
\bibitem{GnatenkoPLA17} Kh.P. Gnatenko, V. M. Tkachuk. Weak equivalence principle in noncommutative phase space and the parameters of noncommutativity. {\it Phys. Lett. A} {\bf 381}, 2463 (2017). doi: 10.1016/j.physleta.2017.05.056
\bibitem{GnatenkoMPLA17} Kh. P. Gnatenko. Kinematic variables in noncommutative phase space and parameters of noncommutativity. {\it Mod. Phys. Lett. A} {\bf32},  1750166 (2017). doi: 10.1142/S0217732317501668
    \end{thebibliography}
    \end{document}